%% file: main.tex
\documentclass[lettersize,journal,comsoc]{IEEEtran}

\usepackage{amsmath,amsfonts}
\usepackage{amsthm}
\usepackage{algorithmic}
\usepackage{algorithm}
\usepackage{bbm}
\usepackage{array}
\usepackage[caption=false,font=normalsize,labelfont=sf,textfont=sf]{subfig}
\usepackage{textcomp}
\usepackage{stfloats}
\usepackage{url}
\usepackage{verbatim}
\usepackage{graphicx}
\usepackage{cite}
\usepackage{bm}
\usepackage{soul,xcolor}
\usepackage{balance}
\usepackage{nicefrac}
\usepackage{multirow}
\usepackage{moresize}

\usepackage{tikz}
\usetikzlibrary{math, calc, positioning, shapes, backgrounds, fit, arrows, decorations.pathreplacing, graphs, spy}
\usepackage{pgfplots}
\pgfplotsset{compat=1.18}
\usepgfplotslibrary{fillbetween}
\usepackage{amsmath}
\pgfplotsset{compat=1.18}

\newcommand{\br}[1]{\left(#1\right)}
\newcommand{\brs}[1]{\left[#1\right]}
\newcommand{\brc}[1]{\left\{#1\right\}}

\newcommand{\hmset}[2]{#1_{m_{\mathcal{#2}}}}

\newcommand{\phix}{\varphi_{\mathcal{X}}}
\newcommand{\phiy}{\varphi_{\mathcal{Y}}}
\newcommand{\phiz}{\varphi_{\mathcal{Z}}}
\newcommand{\phidiff}[1]{\varphi_{\mathcal{X}\setminus \{#1\}}}

\newtheorem{statement}{Statement}

\DeclareMathOperator*{\argmin}{arg\,min}
\newcommand{\arctanh}{\tanh^{-1}}
\input{color_scheme}

\begin{document}
\setstcolor{red}

\title{Generalized LDPC codes with low-complexity decoding and fast convergence}

\author{Dawit Simegn, Dmitry Artemasov, Kirill Andreev, Pavel Rybin, Alexey Frolov
\thanks{Dawit Simegn, Dmitry Artemasov, Kirill Andreev, Pavel Rybin, and Alexey Frolov are with the Center for Next Generation Wireless and IoT (NGW), Skolkovo Institute of Science and Technology, Moscow, Russia (emails: dawit.simegn@skoltech.ru, d.artemasov@skoltech.ru, k.andreev@skoltech.ru, p.rybin@skoltech.ru, al.frolov@skoltech.ru).}
\thanks{The research was carried out at Skolkovo Institute of Science and Technology and supported by the Russian Science Foundation (project no. 23-11-00340), \protect\url{https://rscf.ru/en/project/23-11-00340/}}
}

\maketitle

\begin{abstract}
We consider generalized low-density parity-check (GLDPC) codes with component codes that are duals of Cordaro--Wagner codes. Two efficient decoding algorithms are proposed: one based on Hartmann--Rudolph processing, analogous to Sum-Product decoding, and another based on evaluating two hypotheses per bit, referred to as the Min-Sum decoder. Both algorithms are derived using latent variables and a appropriate message-passing schedule. A quantized, protograph-based density evolution procedure is used to optimize GLDPC codes for Min-Sum decoding. Compared to 5G LDPC codes, the proposed GLDPC codes offer similar performance at 50 iterations and significantly better convergence and performance at 10 iterations.
\end{abstract}


\section{Introduction}
\label{sec:intro}

\IEEEPARstart{G}{eneralized} LDPC codes were first proposed by Tanner in \cite{Tanner81}. Similar to conventional LDPC codes, GLDPC codes can be represented by a sparse bipartite graph; however, the single parity-check (SPC) codes traditionally used at the check nodes are replaced by more powerful linear codes, such as Hamming codes \cite{Zig99}, Bose–Chaudhuri–Hocquenghem (BCH) codes \cite{Boutros1999}, Hadamard codes \cite{Hadamar2007}, and others. In what follows, we refer to these as component codes, and the corresponding nodes in the graph are referred to as constraint nodes (CNs). Numerous works have investigated various aspects of GLDPC codes, including their distance properties \cite{Zig99, Tanner2001}, iterative decoding performance using bit-wise maximum a posteriori (MAP) decoding of component codes \cite{Zig99, Boutros1999, Mitchell2020}, and their error floor behavior \cite{Liva08}. Their suitability for ultra-reliable low-latency communications (URLLC) has also been studied \cite{Mitchell2018}. In addition, we highlight the literature on doubly generalized LDPC codes \cite{DGLCPC2009, DGLCPC2010}, which can be viewed as product-like codes or GLDPC codes with two types of component codes. We also mention works on GLDPC codes with mixed checks \cite{GLDPCmixedBEC2019}, where a small portion of generalized CNs is added to a graph otherwise composed of SPC CNs. 

Despite their strong error-correcting capabilities, GLDPC codes have several drawbacks: (a) the complexity of the decoding operations increases due to the introduction of more sophisticated constraints\footnote{While there are papers on simplified constraint node processing \cite{Hirst2002}, these simplifications typically result in performance degradation.}; (b) the construction lacks flexibility, as it is difficult to vary the length of the component codes, unlike in the case of SPC codes; and (c) the design procedures are often complicated and inaccurate, particularly when based on asymptotic threshold analysis. Regarding the last point, we refer specifically to the binary-input additive white Gaussian noise (AWGN) channel. While several works \cite{Hadamar2007, DGLCPC2009} employ extrinsic information transfer (EXIT)-based analysis, the Gaussian approximation used in such methods is often inaccurate. This issue was addressed in \cite{Chang2025} through the use of a Gaussian mixture model; however, to propagate the density through the constraint node, one still needs to perform computationally intensive Monte Carlo simulations.

In this paper, we investigate a specific construction of GLDPC codes with component codes that are duals of Cordaro--Wagner codes \cite{CW}. The parity-check matrix (PCM) of each such component code consists of distinct non-zero columns of height $2$. To minimize the number of codewords with weight $2$, the distribution of these columns should be as close to uniform as possible. The choice of component codes is motivated by the following considerations: by using PCMs with only two rows, we expect the decoding complexity to remain acceptable even under MAP decoding. Additionally, the length of the component codes can be easily varied. In what follows, we refer to these codes as CW-GLDPC codes. This type of construction has been previously proposed in the literature \cite{usatyuk2021gldpc}. The main objective of this paper is to conduct a fair comparison between CW-GLDPC codes and the LDPC codes used in the 5G standard. To this end, we develop efficient algorithms for both decoding and designing CW-GLDPC codes.

Our main contributions can be summarized as follows. We propose two simple yet effective decoding algorithms for CW-GLDPC codes. The first algorithm leverages Hartmann--Rudolph decoding \cite{HR} for component code processing and similar to the conventional Sum-Product (SP) decoder used in binary LDPC codes. The second algorithm, referred to as the Min-Sum (MS) decoder, is derived by evaluating two competing hypotheses for each bit within the component code. We further describe these algorithms by introducing latent variables and outlining an appropriate message-passing (MP) schedule for their implementation. To design the PCM of CW-LDPC codes we utilize a low-complexity, quantized, protograph-based density evolution (DE) procedure tailored to the proposed MS decoder. Finally, we benchmark CW-LDPC codes against standardized 5G LDPC codes across three representative scenarios characterized by different code rates and block lengths. Our results demonstrate that, under equal computational complexity, CW-GLDPC codes offer comparable performance to 5G LDPC codes after 50 decoding iterations, while significantly outperforming them in the early stages of decoding -- achieving faster convergence and superior performance at 10 iterations.

The paper is structured as follows. Section~\ref{sec:gldpc} describes the construction of our proposed GLDPC codes. In Section~\ref{sec:dec_alg} we present message passing algorithm-based decoders. The methods for constructing and optimizing the parity-check matrices (PCMs) of G-LDPC codes for the proposed decoding algorithms are described in Section~\ref{sec:gldpc_opt}. Section~\ref{sec:num_res} covers simulation results, their analysis, and conclusions.

Scalars are denoted by non-bold letters (e.g., $x$ or $X$), vectors by bold lowercase letters (e.g., $\mathbf{x}$), and matrices by bold uppercase letters (e.g., $\mathbf{X}$). The indicator function of an event $\mathcal{E}$ is denoted by $\mathbbm{1}_{\mathcal{E}}$. For any positive integer $n$, we define $[n] = \{1, \dots, n\}$. Consider a matrix $\mathbf{X}$ of size $m \times n$, and let $\mathcal{I} \subseteq [n]$ be a subset. We denote by $\mathbf{X}_\mathcal{I}$ the submatrix of $\mathbf{X}$ formed by the columns indexed by $\mathcal{I}$.

\section{Code Construction}
\label{sec:gldpc}

\subsection{Component code}\label{sec:cw}

Let us consider the dual of the Cordaro--Wagner code. The parity check matrix of such code consists of non-zero columns of height $2$, e.g.

\[
\mathbf{H}_9 = \left[ \mathbf{h}_{1}, \dots, \mathbf{h}_{9} \right] = \begin{bmatrix}
0 & 1 & 1 & 0 & 1 & 1 & 0 & 1 & 1 \\
1 & 0 & 1 & 1 & 0 & 1 & 1 & 0 & 1 \\
\end{bmatrix}.
\]

Let us introduce the following sets
\begin{flalign*}
\mathcal{X} = & \brc{ i: \; \mathbf{h}_{i} = \mathbf{b}_1 \triangleq \begin{bmatrix} 0 & 1 \end{bmatrix}^T}, \\
\mathcal{Y} = & \brc{ i: \; \mathbf{h}_{i} = \mathbf{b}_2 \triangleq \begin{bmatrix} 1 & 0 \end{bmatrix}^T}, \\
\mathcal{Z} = & \brc{ i: \; \mathbf{h}_{i} = \mathbf{b}_3 \triangleq \begin{bmatrix} 1 & 1 \end{bmatrix}^T}.
\end{flalign*}

Then, for $\mathbf{H}_{9}$ we have
\[
\mathcal{X} = \left\{1, 4, 7 \right\}, \quad
\mathcal{Y} = \left\{2, 5, 8 \right\}, \quad
\mathcal{Z} = \left\{3, 6, 9 \right\}.
\]

Now it is easy to check that the number of codewords of weight $2$ in the code can be calculated as $\binom{|\mathcal{X}|}{2} + \binom{|\mathcal{Y}|}{2} + \binom{|\mathcal{Z}|}{2}$ and thus this quantity is minimized in case when $|\mathcal{X}| \approx |\mathcal{Y}| \approx |\mathcal{Z}|$. For this reason to construct a component code of length $n$ we start from semi-infinite matrix
\[
\mathbf{H}^\infty = \left[\mathbf{b}_1 \: \mathbf{b}_2 \: \mathbf{b}_3 \: \mathbf{b}_1 \: \mathbf{b}_2 \: \mathbf{b}_3 \: \ldots \right]
\]
and choose the first $n$ columns to obtain the component code PCM, i.e.
\[
\mathbf{H}_n = \mathbf{H}^\infty_{[n]}.
\]

The chosen component code has the following properties. First, all symbols are protected by two SPC codes. This property can be seen if we add the sum of rows to the PCM of the component code, e.g.
\[
\mathbf{H}_9 = \begin{bmatrix}
0 & 1 & 1 & 0 & 1 & 1 & 0 & 1 & 1 \\
1 & 0 & 1 & 1 & 0 & 1 & 1 & 0 & 1 \\
1 & 1 & 0 & 1 & 1 & 0 & 1 & 1 & 0 \\
\end{bmatrix}
\]
Second, each codeword $\mathbf{c}$ satisfies one of the following conditions (where $\oplus$ denotes the modulo-two sum):
\begin{equation}\label{eq:cwd_sums}
\bigoplus\limits_{i \in \mathcal{X}}c_{i} = \bigoplus\limits_{i \in \mathcal{Y}}c_{i} = \bigoplus\limits_{i \in \mathcal{Z}}c_{i} = v, \quad v \in\brc{0, 1}.
\end{equation}

\subsection{CW-GLDPC codes}

\begin{figure}
    \centering
    \includegraphics{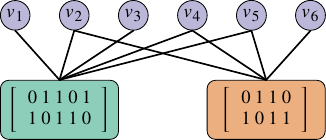}
    \caption{Generalized LDPC structure for six variable nodes and two generalized parity checks having degrees $5$ and $4$.}
    \label{fig:gldpc_structure}
\end{figure}

\begin{figure}
    \centering
    \includegraphics{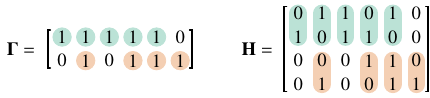}
    \caption{Correspondence between the GLDPC parity check matrix $\mathbf{\Gamma}$ and a parity check matrix $\mathbf{H}$ of the equivalent binary code.}
    \label{fig:gldpc_pcm}
\end{figure}

Now, let us consider the construction of a CW-GLDPC code, starting from a bipartite Tanner graph. The graph's vertex set comprises variable nodes (VNs) $\mathcal{V} = \{v_1, \ldots, v_N\}$, corresponding to code symbols, and constraint nodes $\mathcal{P} = \{p_1, \ldots, p_M\}$, which enforce constraints on these symbols. A VN $v_i$ is connected to a CN $p_j$ if and only if bit $c_i$ participates in constraint $p_j$.

In this construction, the SPC codes typically associated with CNs in LDPC Tanner graphs are replaced by the component codes introduced in Section~\ref{sec:cw}. A CN is satisfied if the bits on its incident edges form a valid codeword of the corresponding component code. A sequence $\mathbf{c} = [c_1, \ldots, c_N]$ is a valid CW-GLDPC codeword if all CNs are satisfied.

An example is shown in Fig.~\ref{fig:gldpc_structure}. The resulting code is linear, and its parity-check matrix $\mathbf{H}$ can be constructed by replacing each $1$ in the graph’s adjacency matrix $\mathbf{\Gamma}$ with a column from the component code’s PCM, and each $0$ with an all-zero column (see Fig.~\ref{fig:gldpc_pcm}).

\section{Decoding algorithms}
\label{sec:dec_alg}

In this section, we consider MP decoding algorithms for CW-GLDPC codes. The VN processing rules are exactly the same as those for conventional LDPC codes; however, the processing rules at the CNs differ significantly, and we describe them in the remainder of the section.

Following the standard turbo decoding strategy (see \cite{KFL, exit}) CN processing includes the bitwise MAP decoding of the corresponding component code and calculation of the extrinsic information. Let us consider the component code $\mathcal{C}$ of length $n$, let $\mathbf{L} = [L_1, \ldots, L_n]$ be the incoming messages, we have the following bitwise MAP rule ($i \in [n]$)
\begin{equation}\label{eq:llr_map}
\hat{L}_{i} =
\ln{\frac{\sigma_{i,0}}{\sigma_{i,1}}} =
\ln{\frac{\sum\limits_{\mathbf{c} \in \mathcal{C}: c_{i}=0} \mu(\mathbf{c})}{\sum\limits_{\mathbf{c} \in \mathcal{C}: c_{i}=1}\mu(\mathbf{c})}},
\end{equation}
where $\mu(\mathbf{c}) = \prod\nolimits_{i=1}^n \mu_i({c}_i)$ and $\mu_i(0)/\mu_i(1) = \exp[L_i]$.

The extrinsic information can be obtained as $\mathbf{E} = \alpha \left( \hat{\mathbf{L}} - \mathbf{L} \right)$, where $\alpha$ is the scaling factor. We note that $\hat{\mathbf{L}}$ can be obtained with use of BCJR algorithm \cite{bcjr} which operates on the trellis representation of the component code. In what follows we do not consider this approach and aim to derive simple closed-form CN processing rules.

\subsection{SP algorithm}
\label{sec:hr_dec}

To reduce complexity, the sums $\sigma_{i,0}$ and $\sigma_{i,1}$ can be calculated using the dual code. This approach is known as the Hartmann--Rudolph decoding \cite{HR}. Let $\mathcal{S} \subseteq [n]$, we define
\begin{equation}\label{eq:phi_def}
\varphi_{\mathcal{S}} = \prod\limits_{i \in \mathcal{S}}\tanh\br{\frac{L_i}{2}}.
\end{equation}

Up to permutation of the columns, the dual code $\mathcal{C}^{\perp}$ is represented by the following four codewords $\mathbf{c}^{\perp}_{i}$, $i=1,\dots,4$.
\begin{flalign*}
\mathbf{c}_1^\perp =& [0\,\dots\,0, \: 0\,\dots\,0, \: 0\,\dots\,0], \\
\mathbf{c}_2^\perp =& [1\,\dots\,1, \: 1\,\dots\,1, \: 0\,\dots\,0], \\
\mathbf{c}_3^\perp =& [1\,\dots\,1, \:  0\,\dots\,0, \:  1\,\dots\,1], \\
\mathbf{c}_4^\perp =& [\underbrace{0\,\dots\,0}_{\mathcal{X}}, \:  \underbrace{1\,\dots\,1}_{\mathcal{Y}}, \:  \underbrace{1\,\dots\,1}_{\mathcal{Z}}].
\end{flalign*}
According to the Hartmann--Rudolph decoding rule, we can write ($s \in \left\{0, 1 \right\}$)
\[
\sigma_{i,s} = \gamma \sum\limits_{\mathbf{c}^{\perp} \in \mathcal{C}^{\perp}}\br{\prod\limits_{j = 1}^{n}\varphi_{\{j\}}^{c^{\perp}_{j}} + \br{-1}^{s}\prod\limits_{j=1}^{n}\varphi_{\{j\}}^{\br{c^{\perp}_{j} \oplus \mathbbm{1}_{i = j}}}},
\]
where $\gamma$ is some value that does not depend on $s$.

Then, if $i \in \mathcal{X}$, using the four codewords of the dual code introduced above, we obtain:
\begin{flalign*}
\frac{1}{\gamma} \sigma_{i,0} = 1 + \varphi_{\{i\}} &+ \phix \phiy + \phiy \phidiff{i} \\
& + \phix \phiz + \phiz \phidiff{i} \\
& + \phiy \phiz + \phiz \phiy \varphi_{\{i\}}=
\end{flalign*}
\[
\br{1 + \varphi_{\{i\}}}\br{1 + \phiy \phidiff{i}  + \phiz \phidiff{i} + \phiz \phiy},
\]
where the last equality holds from $\phix = \varphi_{\{i\}} \phidiff{i}$. Having a similar derivation for $\sigma_{i,1}$, we get
\[
\hat{L}_{i} = \ln{\frac{\br{1 + \varphi_{\{i\}}}\br{ 1 + \phidiff{i} \phiy + \phidiff{i} \phiz + \phiy \phiz}}{\br{1 - \varphi_{\{i\}}} \br{1 - \phidiff{i} \phiy - \phidiff{i} \phiz + \phiy \phiz }}}.
\]

Using the fact that $L_i = \ln[ (1 + \varphi_{\{i\}}) / (1 - \varphi_{\{i\}})]$ and the following equality
$$
\ln\frac{1 + ab + ac + bc}{1 - ab - ac + bc}= \ln\frac{1 + a\frac{b + c}{1 + bc}}{1 - a\frac{b + c}{1 + bc}} = 2 \arctanh\br{a\frac{b + c}{1 + bc}},
$$
we get
\begin{flalign}
\hat{L}_{i} & = L_{i} + 2\arctanh{\brs{\phidiff{i} \cdot \frac{\phiy+\phiz}{1+\phiy \phiz} }} \label{eq:hr_update_final}\\
& = L_{i} + 2\arctanh\brs{  \phidiff{i}\cdot\tanh\br{ \arctanh\phiy + \arctanh\phiz}}. \nonumber
\end{flalign}

\subsection{MS algorithm}
\label{sec:ms_dec}
The expression~\eqref{eq:llr_map} can be approximated as follows:
\[
\hat{L}_{i} \approx 
\ln{\frac{\max\limits_{\mathbf{c} \in \mathcal{C}: c_{i}=0} \mu(\mathbf{c})}{\max\limits_{\mathbf{c} \in \mathcal{C}: c_{i}=1} \mu(\mathbf{c})}}.
\]

Let $\mathbf{c}^{(i)}_0$ and $\mathbf{c}^{(i)}_1$ be the codewords that maximize the numerator and denominator, respectively. These are the closest codewords with $0$ and $1$ in the $i$-th position, representing two competing hypotheses. Given the simplicity of the component code, such codewords can be constructed explicitly.

Taking into account~\eqref{eq:cwd_sums}, we consider four codewords $\mathbf{c}^{(i)}_{s, v}$ -- the most probable codewords with $c_{i} = s \in \{0, 1\}$ and fixed $v \in \{0, 1\}$. We can now write:
\begin{equation}\label{eq:llr_ms}
\hat{L}
\approx \ln{\frac{\max\left\{ \mu\left( \mathbf{c}^{(i)}_{0, 0} \right), \mu\left( \mathbf{c}^{(i)}_{0, 1} \right) \right\}}{\max\left\{ \mu\left( \mathbf{c}^{(i)}_{1, 0} \right), \mu\left( \mathbf{c}^{(i)}_{1, 1} \right) \right\}}}.
\end{equation}

The idea is similar to SPC code decoding: we need to consider only the positions with the smallest LLR magnitudes in each of the sets $\mathcal{X}$, $\mathcal{Y}$, and $\mathcal{Z}$. Without loss of generality, we assume that $i \in \mathcal{X}$. We introduce the index $m_\mathcal{X}$ and the LLR value $\tilde{L}_\mathcal{X}^{(i)}$ as follows:
\begin{equation}\label{eq:ms_terms}
m_\mathcal{X} \triangleq  \argmin\limits_{j \in \mathcal{X} \setminus \left\{ i \right\}} \left| L_{j} \right|,
\quad
\tilde{L}_\mathcal{X}^{(i)} =  \left( -1 \right)^{\bigoplus\limits_{j \in \mathcal{X} \setminus \{ i \}} s_{j}} \left| \hmset{L}{X} \right|,
\end{equation}
where $s_j = \mathbbm{1}_{L_j < 0}$. These quantities can be defined similarly for the sets $\mathcal{Y}$ and $\mathcal{Z}$.

Let $\hmset{h}{X}$, $\hmset{h}{Y}$, and $\hmset{h}{Z}$ be the values of hypothesis $\mathbf{c}^{(i)}_{0, 0}$ at positions $m_{\mathcal{X}}$, $m_{\mathcal{Y}}$, and $m_{\mathcal{Z}}$, respectively. Clearly,
\[
\hmset{h}{X} = \bigoplus\limits_{j \in \mathcal{X} \setminus \{ i, m_{\mathcal{X}} \}} s_{j}, \:
\hmset{h}{Y} = \bigoplus\limits_{j \in \mathcal{Y} \setminus \{ m_{\mathcal{Y}} \}} s_{j}, \:
\hmset{h}{Z} = \bigoplus\limits_{j \in \mathcal{Z} \setminus \{ m_{\mathcal{Z}} \}} s_{j}.
\]

We have
\begin{flalign*}
\mu\left( \mathbf{c}^{(i)}_{0, 0} \right) &\propto \mu_i(0) \mu_{m_{\mathcal{X}}} (\hmset{h}{X}) \mu_{m_{\mathcal{Y}}} (\hmset{h}{Y}) \mu_{m_{\mathcal{Z}}} (\hmset{h}{Z}), \\
\mu\left( \mathbf{c}^{(i)}_{0, 1} \right) &\propto \mu_i(0) \mu_{m_{\mathcal{X}}} (\hmset{h}{X} \oplus 1) \mu_{m_{\mathcal{Y}}} (\hmset{h}{Y} \oplus 1) \mu_{m_{\mathcal{Z}}} (\hmset{h}{Z} \oplus 1), \\
\mu\left( \mathbf{c}^{(i)}_{1, 0} \right) &\propto \mu_i(1) \mu_{m_{\mathcal{X}}} (\hmset{h}{X} \oplus 1) \mu_{m_{\mathcal{Y}}} (\hmset{h}{Y}) \mu_{m_{\mathcal{Z}}} (\hmset{h}{Z}), \\
\mu\left( \mathbf{c}^{(i)}_{1, 1} \right) &\propto \mu_i(1) \mu_{m_{\mathcal{X}}} (\hmset{h}{X}) \mu_{m_{\mathcal{Y}}} (\hmset{h}{Y} \oplus 1) \mu_{m_{\mathcal{Z}}} (\hmset{h}{Z} \oplus 1),
\end{flalign*}
where we use the symbol $\propto$ instead of $=$ since the values are to be multiplied by a common constant\footnote{This constant reflects the fact that differences occur in only four positions, while the remaining codeword positions are identical.}.

Next, making simple transformations we can write:
\[
\hat{L}_{i} \approx L_{i} + E_i,
\]
where
\begin{equation}
E_i = \max{\left\{
\tilde{L}_\mathcal{X}^{(i)} + \tilde{L}_\mathcal{Y} + \tilde{L}_\mathcal{Z},
0
\right\}} - \max{ \left\{
\tilde{L}_\mathcal{Y} + \tilde{L}_\mathcal{Z}, \tilde{L}_\mathcal{X}^{(i)}
\right\} } \label{eq:llr_ms_final},
\end{equation}
and the superscript $(i)$ is omitted for $\tilde{L}_\mathcal{Y}$ and $\tilde{L}_\mathcal{Z}$, since $i \in \mathcal{X}$. Note that the update rules for $i \in \mathcal{Y}$ and $i \in \mathcal{Z}$ take a similar form.

\subsection{Decoding with use of latent variables}

\begin{figure}
    \centering
    \includegraphics{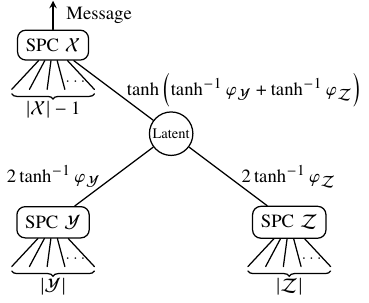}
    \caption{Graphical representation of the rule~\eqref{eq:hr_update_final}.}
    \label{fig:gldpc_latent}
\end{figure}

Let us now examine the expression~\eqref{eq:hr_update_final} in more detail and represent it using a factor graph~\cite{KFL}. This rule can be interpreted as the result of introducing a latent variable into the component code's PCM, as shown in~\eqref{eq:latent_pcm}: 

\begin{equation}\label{eq:latent_pcm}
\mathbf{H} = 
\underbrace{\left[\begin{array}{c} 1\: 1\: \cdots \: 1 \\ \\ \\ \end{array}\right.}_{\mathcal{X}} \:
\underbrace{\left.\begin{array}{c} \\ 1\: 1\: \cdots \: 1 \\ \\ \end{array}\right.}_{\mathcal{Y}} \:
\underbrace{\left.\begin{array}{c} \\ \\ 1\: 1\: \cdots \: 1 \end{array}\right.}_{\mathcal{Z}} \:
\underbrace{\left.\begin{array}{c} 1 \\ 1 \\ 1 \end{array}\right]}_{\mathcal{L}}
\end{equation}

A natural question arises: what happens if we apply the standard min-sum (MS) decoding algorithm to the matrix~\eqref{eq:latent_pcm}? From Fig.~\ref{fig:gldpc_latent}, we derive the following update rule:
\begin{equation}\label{eq:ms_latent}
E'_i = (-1)^{s\left(\tilde{L}_\mathcal{X}^{(i)}\right)} (-1)^{s\left(\tilde{L}_\mathcal{Y} + \tilde{L}_\mathcal{Z}\right)} \min\{|\tilde{L}_\mathcal{X}^{(i)}|, |\tilde{L}_\mathcal{Y} + \tilde{L}_\mathcal{Z}|\},
\end{equation}
where $s(L) = \mathbbm{1}_{L < 0}$.

By exhaustively considering all possible sign combinations and magnitude relationships, one can verify the following result:

\begin{statement}
The extrinsic values computed by~\eqref{eq:llr_ms_final} and~\eqref{eq:ms_latent} are identical, i.e., $E_i = E'_i$.
\end{statement}

Thus, the CN update rules can be implemented using the standard SP or MS algorithms applied to a factor graph with a latent variable. It is important to note that the scheduling (i.e., the order of updates) also plays a critical role in the overall decoding performance (see the numerical results).

\section{Parity check matrix optimization}
\label{sec:gldpc_opt}
To optimize the CW-GLDPC PCM for the MS decoding algorithm~\eqref{eq:ms_latent}, we use a DE framework. The decoding performance of a given PCM is first evaluated via DE, followed by an optimization procedure.

We perform quantized DE on a \emph{protograph} structure~\cite{Liva2007proto}, which defines a constrained subset of the GLDPC ensemble by specifying degree distributions and edge connections. The only modification to the standard DE procedure is the CN update rule~\eqref{eq:ms_latent}, implemented as a sequence of classical MS updates.

Optimization is carried out using a genetic algorithm: the best-performing PCMs are slightly perturbed to produce the next generation. The signal-to-noise ratio (SNR) is gradually decreased in small steps, guiding the search toward an optimal base graph that meets the desired error threshold.

\section{Numerical results}
\label{sec:num_res}

In this section, we evaluate the performance of CW-GLDPC codes and compare them to 5G-LDPC codes using various MP decoding algorithms. The primary performance metric is block error rate (BLER) as a function of $E_{s}/N_{0}$. Following standard 5G NR practices, we use quadrature phase shift keying (QPSK) modulation over an AWGN channel.

\begin{figure}[!htbp]
\nocite{PolyanskiyPoorVerdu}
\includegraphics{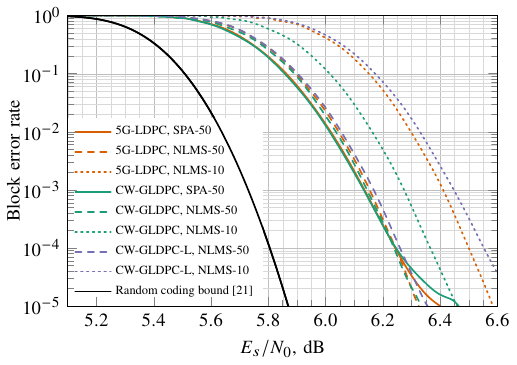}
\caption{Simulation results for scenario A.\label{fig:plot_R88}}
\end{figure}

\begin{figure}[!htbp]
\includegraphics{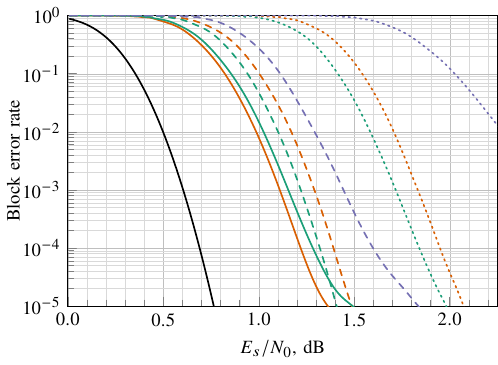}
\caption{Simulation results for scenario B.\label{fig:plot_R50}}
\end{figure}

\begin{figure}[!htbp]
\includegraphics{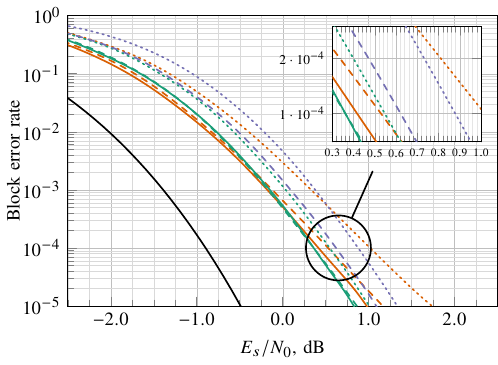}
\caption{Simulation results for scenario C.\label{fig:plot_R25}}
\end{figure}

We consider three representative scenarios A, B and C. Code parameters and performance gains of CW-GLDPC over 5G-LDPC codes are summarized in Table~\ref{tab:sim_scenario_gain}. Simulation results for scenarios A, B, and C are shown in Fig.~\ref{fig:plot_R88}, Fig.~\ref{fig:plot_R50}, and Fig.~\ref{fig:plot_R25}, respectively (the legend in Fig.~\ref{fig:plot_R88} is common for all figures).

Although performance comparisons are often based on the SP algorithm with 50 iterations, practical solutions typically employ MS-like decoders. Therefore, we also evaluate the normalized layered min-sum\footnote{A layer-wise scheduled MS decoding algorithm with scaling factor $\alpha$ (we fix $\alpha = 0.75$ for all cases).} (NLMS) algorithm with $50$ and $10$ iterations. Note that the CW-GLDPC codes considered here were optimized specifically for 50 NLMS iterations. We also note that, for each figure, the same code is used to evaluate all curves of the same color.

As shown in Table~\ref{tab:sim_scenario_gain}, CW-GLDPC codes outperform 5G-LDPC codes under NLMS decoding, particularly with 10 iterations in scenarios A and B. This performance gain is even more pronounced in scenario C. These results highlight the superior convergence behavior of CW-GLDPC codes under the proposed MS decoding approach -- an important advantage for future communication system design.

Additionally, we evaluate a variant of CW-GLDPC codes represented via latent variables (denoted CW-GLDPC-L; see Fig.~\ref{fig:gldpc_latent}), decoded as an LDPC code with punctured latent symbols. As you can see, this leads to significant performance degradation. In other words, the CW-GLDPC structure inherently introduces punctured symbols, and the proposed decoding rules effectively schedule their recovery.

\begin{table*}[t]
\centering
\caption{Parameters of the 5G LDPC and CW-GLDPC codes and performance gains for different scenarios}
\label{tab:sim_scenario_gain}
\begin{tabular}{|c|ccccccc|ccc|}
\hline
\multirow{2}{*}{Scenario} &
  \multicolumn{7}{c|}{Code parameters} &
  \multicolumn{3}{c|}{Gain at BLER=$10^{-4}$, dB} \\ \cline{2-11} 
 &
  \multicolumn{1}{c|}{Code} &
  \multicolumn{1}{c|}{{[}N, K{]}} &
  \multicolumn{1}{c|}{R} &
  \multicolumn{1}{c|}{\begin{tabular}[c]{@{}c@{}}Base\\ matrix\end{tabular}} &
  \multicolumn{1}{c|}{\begin{tabular}[c]{@{}c@{}}Lifting\\ factor\end{tabular}} &
  \multicolumn{1}{c|}{\begin{tabular}[c]{@{}c@{}}Punctured\\ symbols\end{tabular}} &
  \begin{tabular}[c]{@{}c@{}}Av. num. of \\ ones in column\end{tabular} &
  \multicolumn{1}{c|}{SP-50} &
  \multicolumn{1}{c|}{NLMS-50} &
  NLMS-10 \\ \hline
\multirow{2}{*}{A} &
  \multicolumn{1}{c|}{5G LDPC} &
  \multicolumn{1}{c|}{\multirow{2}{*}{{[}6000, 5280{]}}} &
  \multicolumn{1}{c|}{\multirow{2}{*}{0.88}} &
  \multicolumn{1}{c|}{$5\times27$} &
  \multicolumn{1}{c|}{240} &
  \multicolumn{1}{c|}{$2\times240$} &
  2.93 &
  \multicolumn{1}{c|}{\multirow{2}{*}{-0.003}} &
  \multicolumn{1}{c|}{\multirow{2}{*}{0.006}} &
  \multirow{2}{*}{0.134} \\ \cline{2-2} \cline{5-8}
 &
  \multicolumn{1}{c|}{CW-GLDPC} &
  \multicolumn{1}{c|}{} &
  \multicolumn{1}{c|}{} &
  \multicolumn{1}{c|}{$7\times102$} &
  \multicolumn{1}{c|}{60} &
  \multicolumn{1}{c|}{$2\times60$} &
  2.85 &
  \multicolumn{1}{c|}{} &
  \multicolumn{1}{c|}{} &
   \\ \hline
\multirow{2}{*}{B} &
  \multicolumn{1}{c|}{5G LDPC} &
  \multicolumn{1}{c|}{\multirow{2}{*}{{[}5760, 2880{]}}} &
  \multicolumn{1}{c|}{\multirow{2}{*}{0.50}} &
  \multicolumn{1}{c|}{$12\times22$} &
  \multicolumn{1}{c|}{288} &
  \multicolumn{1}{c|}{$2\times288$} &
  3.50 &
  \multicolumn{1}{c|}{\multirow{2}{*}{-0.064}} &
  \multicolumn{1}{c|}{\multirow{2}{*}{0.071}} &
  \multirow{2}{*}{0.109} \\ \cline{2-2} \cline{5-8}
 &
  \multicolumn{1}{c|}{CW-GLDPC} &
  \multicolumn{1}{c|}{} &
  \multicolumn{1}{c|}{} &
  \multicolumn{1}{c|}{$12\times48$} &
  \multicolumn{1}{c|}{120} &
  \multicolumn{1}{c|}{$2\times120$} &
  2.77 &
  \multicolumn{1}{c|}{} &
  \multicolumn{1}{c|}{} &
   \\ \hline
\multirow{2}{*}{C} &
  \multicolumn{1}{c|}{5G LDPC} &
  \multicolumn{1}{c|}{{[}264, 66{]}} &
  \multicolumn{1}{c|}{\multirow{2}{*}{0.25}} &
  \multicolumn{1}{c|}{$20\times26$} &
  \multicolumn{1}{c|}{11} &
  \multicolumn{1}{c|}{$2\times11$} &
  3.39 &
  \multicolumn{1}{c|}{\multirow{2}{*}{0.056}} &
  \multicolumn{1}{c|}{\multirow{2}{*}{0.169}} &
  \multirow{2}{*}{0.470} \\ \cline{2-3} \cline{5-8}
 &
  \multicolumn{1}{c|}{CW-GLDPC} &
  \multicolumn{1}{c|}{{[}256, 64{]}} &
  \multicolumn{1}{c|}{} &
  \multicolumn{1}{c|}{$6\times16$} &
  \multicolumn{1}{c|}{16} &
  \multicolumn{1}{c|}{-} &
  2.63 &
  \multicolumn{1}{c|}{} &
  \multicolumn{1}{c|}{} &
   \\ \hline
\end{tabular}
\end{table*}

\balance

\bibliographystyle{IEEEtran}
\bibliography{main}

\vfill

\end{document}

%% file: color_scheme.tex
\ifdefined\DARK
    \def\PgfColorA{255,255,255}
\else
    \def\PgfColorA{0,0,0}
\fi

\def\PgfColorB{27,158,119}
\def\PgfColorC{217,95,2}
\def\PgfColorD{117,112,179}
\def\PgfColorE{231,41,138}

\definecolor{MainColorA}{RGB}{\PgfColorA}
\definecolor{MainColorB}{RGB}{\PgfColorB}
\definecolor{MainColorC}{RGB}{\PgfColorC}
\definecolor{MainColorD}{RGB}{\PgfColorD}
\definecolor{MainColorE}{RGB}{\PgfColorE}